\documentclass[11pt,twoside]{article}

\usepackage{asp2014}
\usepackage{hyperref}
\usepackage{cleveref}

\aspSuppressVolSlug
\resetcounters

\begin{document}

\title{Astrophysical code migration into Exascale Era}

\author{D. Goz,$^1$ S. Bertocco,$^1$ L. Tornatore,$^1$ and G. Taffoni,$^1$}
\affil{$^1$INAF - Osservatorio Astronomico di Trieste - via Tiepolo 11, 34131 Trieste Italy; \email{david.goz@inaf.it}}

\paperauthor{D. Goz}{david.goz@inaf.it}{orcid.org/0000-0001-9808-2283}{INAF}{Osservatorio Astronomico di Trieste}{Trieste}{Trieste}{34131}{Italy}


\begin{abstract}
The \texttt{ExaNeSt} and \texttt{EuroExa} H2020 EU-funded projects aim to design and develop an exascale ready computing platform prototype based on low-energy-consumption ARM64 cores and FPGA accelerators. We participate in the application-driven design of the hardware solutions and  prototype validation. To carry on this work we are using, among others, \texttt{Hy-\emph{N}body}, a state-of-the-art direct \emph{N}-body code. Core algorithms of \texttt{Hy-\emph{N}body} have been improved in such a way to increasingly fit them to the exascale target platform.
Waiting for the ExaNest prototype release, we are performing tests and code tuning operations on an ARM64 SoC facility: a \texttt{SLURM} managed HPC cluster based on 64-bit ARMv8 Cortex-A72/Cortex-A53 core design and powered by a Mali-T864 embedded GPU. In parallel, we are porting a kernel of \texttt{Hy-\emph{N}body} on FPGA aiming to test and compare the performance-per-watt of our algorithms on different platforms.
In this paper we describe how we re-engineered the application and we show first results on ARM SoC.

\end{abstract}
\section{Introduction}
The current market offers low-power micro-processor hardware solutions integrating enough transistors to include an on-chip floating-point unit capable of running typical HPC (High Performance Computing) applications. They are less expensive and more power-efficient than standard HPC devices. For this reason SoC (System on Chip) solutions are a possible approach to actually reduce the costs of HPC in terms of time and power consumption and this becomes extremely important when designing the new generation of
HPC supercomputer, the Exascale platforms.
The \texttt{ExaNest} H2020 project \citep{ExaNeSt} aims at the design and 
development of an exascale-class prototype computing system built upon power-efficient hardware able to execute real-world applications coming from a wide range of scientific and industrial domains, including also HPC for astrophysics \citep{8049832}. 
The ExaNeSt basic compute unit consists of low-energy-consumption ARM CPUs, FPGAs and low-latency interconnects \citep{KATEVENIS201858}.

Programmers will have to re-engineer their applications in order to fully exploit this new exascale platform based on heterogeneous hardware. We studied whether a direct $N$-body code for real scientific production may benefit from embedded GPUs given that the powerful high-end GPUs already have demonstrated to provide tremendous performance benefit for $N$-body code. 
To the best of our knowledge, this is the first work to implement such algorithm on embedded GPUs and to compare results with multi-core solutions on a SoC implementation.

\section{Code implementation}
\label{label:code_implementation}
\texttt{Hy-\emph{N}body} is a direct \emph{N}-body code that relies on the Hermite 6th order time integrator and that has been conceived to exploit hybrid hardware. The code is derived from \texttt{HiGPUs} \citep{HiGPUs, HiGPUs2, HiGPUs3}, which has been widely used for simulations of star clusters with up to $\sim$ 8 million bodies \citep{Spera1,Spera2}, and of galaxy mergers \citep{Spera3}.
The kernels of \texttt{Hy-\emph{N}body} have been developed with OpenCL in order to write efficient code for hybrid (CPU/GPU/FPGA) architecture.
Kernels have been optimized using \emph{(i) vectorization}, to increase the number of operations per cycle, and exploiting the \emph{(ii) local memory} of the device, to reduce the latency of data transactions.
The OpenCL host code is parallelized with hybrid MPI+OpenMP programming.
A one-to-one correspondence between MPI processes and computational nodes is established and each MPI process manages all the OpenCL-compliant devices of the same type available per node. Inside of each shared-memory computational node, parallelization is achieved by means of OpenMP environment.

The Hermite 6th order integration schema requires double precision (DP) arithmetic in the evaluation of inter-particles distance and acceleration in order to minimize the round-off error.
Full IEEE-compliant DP-arithmetic is efficient in available CPUs and GPGPUs, but it is still extremely resource-eager and performance-poor in other accelerators like embedded GPUs or FPGAs.
The extended-precision (EX) numeric type is a valuable alternative in porting our application on devices not specifically designed for scientific calculations, such as embedded GPUs or FPGAs. 
We implemented in \texttt{Hy-\emph{N}body} the EX-arithmetic as proposed by \cite{EX}.

On SoC the memory is shared between CPU and GPU
so, using local memory as a cache with associated barrier synchronization can waste both performance and power. For this reason, we implemented a specific embedded-GPU-optimized version of all kernels of \texttt{Hy-\emph{N}body}.

\section{Testbed description}
We deployed a cluster based on heterogeneous hardware (CPU+GPU) to validate and test the {\texttt{Hy-\emph{N}body}} code.
Each computational node is a Rockchip Firefly-RK3399 single board computer. 
It is a six core 64-bit High-Performance Platform, based on SoC with the ARM big.LITTLE architecture.
The main characteristics of this cluster, named \texttt{INCAS}\footnote{\textbf{IN}tensive \textbf{C}lustered \textbf{A}rm-\textbf{Soc}}, are listed in Table~\ref{table:INCAS}, while full details are in \cite{Bertocco_INAF_report_1}.

\begin{table}
\begin{center}
\begin{tabular}{|c|c|}
\hline
Cluster name & \texttt{INCAS}\\
\hline
Nodes available & 8\\
\hline
SoC         & Rockchip RK3399 (28nm HKMG Process)\\
\hline
CPU   & Six-Core ARM 64-bit processor \\
& (Dual-Core Cortex-A72 and Quad-Core Cortex-A53)\\
\hline
GPU & ARM Mali-T864 MP4 Quad-Core GPU\\
\hline
Ram memory & 4GB Dual-Channel DDR3 (per node)\\
\hline
Network & 1000Mbps Ethernet\\
\hline
Power & DC12V - 2A (per node)\\
\hline
Operating System & Ubuntu version 16.04 LTS\\
\hline
Compiler & gcc version 7.3.0\\
\hline
MPI      & OpenMPI version 3.0.1\\
\hline
OpenCL   & OpenCL 2.2\\
\hline
Job scheduler & SLURM version 17.11\\
\hline
\end{tabular}
\end{center}
\caption{The main characteristics of our cluster used to test the \texttt{Hy-\emph{N}body} code.}
\label{table:INCAS}
\end{table}

\section{Performance results}

We just focused on the most computationally demanding kernel of the Hermite 6th order algorithm (with $N$ bodies the kernel has $O(N^{2})$ computational cost) and compared the performances on ARM CPUs.
\begin{figure}[h]
\centering\includegraphics[width=\linewidth]{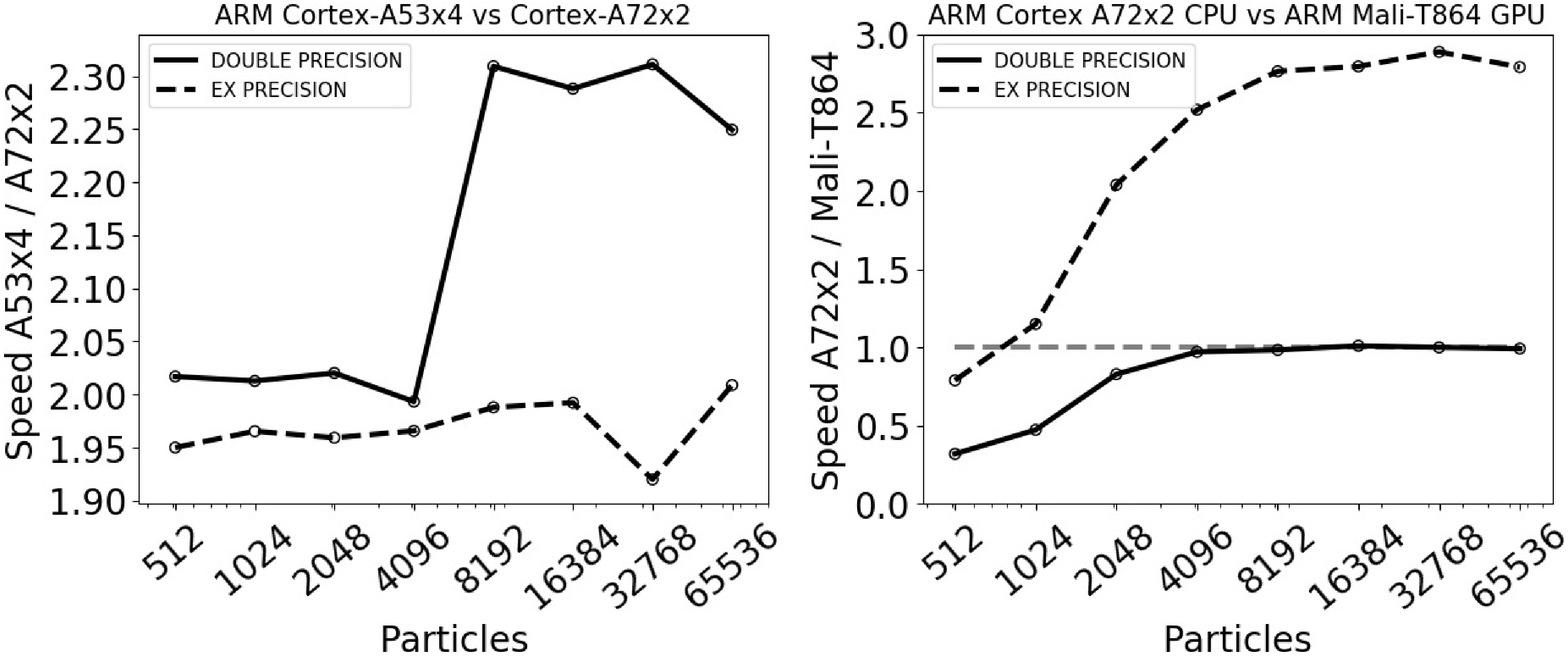}
\caption{Left panel: speed comparison between ARM Cortex-A53x4 and Cortex-A72x2 CPUs for both DP-arithmetic (continuous line) and EX-arithmetic (dashed line) as a function of the number of particles. Right panel: comparison of the time to solution between ARM Cortex-A72x2 CPU and Mali-T864 GPU for both DP-arithmetic (continuous line) and EX-arithmetic (dashed line) as a function of the number of particles.}
\label{fig:host_CPUs_comparison_time}
\end{figure}
Left panel of Figure~\ref{fig:host_CPUs_comparison_time} shows the ratio of the best running time achieved by the CPUs as a function of the number of particles for both arithmetic. ARM Cortex-A72 with two cores is faster than Cortex-A53 with four cores by approximately a factor of two.

High-end GPGPUs have already proved to speedup the solution of the direct $N$-body problem. In this work we aim to evaluate the performance of low-power embedded ARM GPU. We studied the best running time on ARM Cortex-A72x2 as the ratio over the best execution time taken by our ARM-optimized GPU implementation, as shown in the right panel of Figure~\ref{fig:host_CPUs_comparison_time}.
The ARM-optimized implementation is as fast as the dual-core implementation on the ARM Cortex-A72x2 using DP-arithmetic, as long as the GPU is kept fed with enough particles, while is almost three times faster using EX-precision.

\section{Future development}
ExaNeSt project is facing, among others, the challenge of the sustainable power consumption focusing on efficient hardware acceleration. For this reason, we are planning also to quantitatively measure the impact of our algorithms on energy consumption on SoC, shedding some light on their suitability for exascale applications.
The findings from this research activity on ARM SoC are fundamental in order to also enhance our capabilities to exploit FPGAs for HPC, which in comparison to both CPUs and GPUs provide higher throughput-per-watt. 

\section{Conclusions}
In light of our findings, embedded GPUs appear to be attractive from a performance perspective as soon as their double-precision compute capability increases. However, we demonstrated that the extended-precision approach can be a solution to supply enough power to execute scientific computation and benefit at maximum of the SoC devices. 

SoC technology will play a fundamental role on future Exascale heterogeneous platforms that will involve millions of specialized parallel compute units.
Programmers will have to re-design their codes in order to fully exploit embedded accelerators, because of restricted hardware features compared to high-end GPGPUs.

\acknowledgements 

This work was carried out within the ExaNeSt (FET-HPC) project (grant no. 671553) and the ASTERICS project (grant no. 653477), funded by  the European Union's Horizon 2020 research and innovation program.

\bibliography{P12-5}  

\end{document}